\documentclass[prl,twocolumn]{revtex4}
\usepackage{epsfig}
\usepackage{color}
\usepackage{braket}
\usepackage{graphicx}
\usepackage[bookmarks=false,hyperfootnotes=false]{hyperref}
\hypersetup{
			colorlinks=true,
			linkcolor=blue,
			filecolor=blue
			anchorcolor=blue,
			citecolor=blue,
			urlcolor=blue
}
\usepackage{amssymb,amsfonts,amsmath}
\usepackage{tikz}
\definecolor{green}{rgb}{0.2, 0.8, 0.2}

\begin{document}

\title{Discrete Eulerian model for population genetics and dynamics under flow}

\author{Giorgia Guccione}

\affiliation{Department of Applied Physics, Eindhoven University of Technology,5600 MB Eindhoven, The Netherlands 
and Department of Physics and INFN, University of Tor Vergata, Via della Ricerca Scientifica 1, I-00133 Rome, Italy}

\author{Roberto Benzi}
\affiliation{Department of Physics and INFN, University of Tor Vergata, Via della Ricerca Scientifica 1, I-00133 Rome, Italy}

\author{Abigail Plummer}
\affiliation{Department of Physics, Harvard University, 17 Oxford Street, Cambridge, Massachusetts 02138, USA}

\author{Federico Toschi}
\affiliation{Department of Applied Physics, Department of Mathematics and Computer Science, 
Eindhoven University of Technology, 5600 MB Eindhoven, The Netherlands and CNR-IAC, Via  dei Taurini 19, I-00185 Rome, Italy}

\date{December 3, 2019}

\begin{abstract}
Marine species reproduce and compete while being advected by turbulent flows. It is largely unknown, both theoretically and experimentally, how population dynamics and genetics are changed by the presence of fluid flows. Discrete agent-based simulations in continuous space allow for accurate treatment of advection and number fluctuations, but can be computationally expensive for even modest organism densities. In this report, we propose an algorithm to overcome some of these challenges. We first provide a thorough validation of the algorithm in one and two dimensions without flow. Next, we focus on the case of weakly compressible flows in two dimensions. This models organisms such as phytoplankton living at a specific depth in the three-dimensional, incompressible ocean experiencing upwelling and/or downwelling events. We show that organisms born at sources in a two-dimensional time-independent flow experience an increase in fixation probability.
\end{abstract}

%\pacs{}

\maketitle
%%%%%%%%%%%%%%%%%%%%%%%%%%%%%%%%
%
% INTRODUCTION
%
%%%%%%%%%%%%%%%%%%%%%%%%%%%%%%%%
\section{I. Introduction}

Marine plankton account for roughly half of the total biological production on Earth; they are responsible for most of the transfer of carbon dioxide from the atmosphere to the ocean ~\cite{guasto2012fluid,field1998primary}. 
Planktonic organisms are  an essential part of the global carbon cycle, and even small changes in their productivity or in the relative abundance of the thousands of species could have a substantial influence on climate change ~\cite{field1998primary}. It is important to understand the variation in physical factors that a population can withstand and how it can continue to thrive in high Reynolds number fluid environments in order to support our oceanic ecosystem  ~\cite{perlekar2010population}.

Microorganism populations are carried along the uppermost layer (euphotic zone $\sim$100 m) of the ocean ~\cite{d2010fluid}. 

The euphotic zone is characterized by a low quantity of nutrients due to consumption by phytoplankton. 
Periodic events, such as upwelling and downwelling currents, supply nutrients to the upper water column. The aforementioned mechanisms can trigger the processes of water exchange in the mixed layer of the ocean. The upwelling current leads to a rising up of deep water, where a rich concentration of nutrients resides.
Passively traveling organisms, transported by the ocean circulation, experience compressible turbulence  ~\cite{benzi2012population}  from the convergence or the divergence of water masses.

The study of genetic variation within a population deals with the biological differences affecting reproduction, feeding strategies, disease resistance, and many other factors. 
Well-adapted individuals with inherited favorable characteristics may survive and grow faster than others, passing on the genes that make them successful; such organisms have a selective advantage. 

If we consider two species, one with a selective advantage and one without, called $A$ and $B$, respectively, 
it is not possible to determinate {\it a priori} which one of the two will be the dominant one in the presence of turbulence.
However it is possible to calculate a probability.
In the absence of advection, Kimura ~\cite{kimura1962probability} derived a theoretical prediction for the fixation probability of one of the species for the well-mixed case,
\begin{equation}
\label{eq:kim}
P_{\text{fix}} = \frac{1-e ^{-s N f}}{1- e ^{- s N}}. 
\end{equation}
 
This formula describes the fixation probability for a species with selective advantage $s$ in a population of size $N$ that makes up an initial fraction $f$ of all organisms, neglecting any space dependency. This result can be applied to a spatially extended population with simple migration patterns, such as diffusion \cite{maruyama1974simple}.
Several stochastic models for genetic evolution have been developed. 
Among these, we mention the Moran model \cite{moran1958random}, a simple approach that takes into consideration the selection of the organisms and the genetic drift, and the Stepping Stone model  \cite{korolev2010genetic},  an extension of the previous one including the migration and reproduction of the individuals. These models share many similarities with the ones used to investigate
nonequilibrium phase transitions (see \cite{hinrichsen2000non, dornic2005integration}). The aforementioned models are tailored for lattice rules and do not allow straightforward generalization to take into account an external velocity advection. In  \cite{pigolotti2013growth}, an alternative method has been introduced: each individual is advected by the external velocity and diffusion is implemented by a stochastic noise, while death and reproduction processes are implemented using an interaction distance $\delta$. This requires an extra computational cost to evaluate the individual numbers in each virtual deme of size $\delta$. This method has been recently used in ~\cite{plummer2018fixation}, where the competition between two different species, distributed in continuous space, and under the effect of a compressible flow  is examined through an agent-based model. It has been shown that a turbulent flow can dramatically change outcomes and, in particular, it can reduce the effect of selective advantage on fixation probabilities ~\cite{plummer2018fixation,pigolotti2012population,benzi2009fisher}. 

In this paper, we propose a computational approach which merges the accuracy of working in continuous space with the efficiency of working on a lattice.
We assume a uniform lattice of spacing $\Delta x$ with each site occupied by $N_j$ individuals. At each
time step, we redistribute the $N_j$ individuals on a domain $(1+a)\Delta x^d$, where $d$ is the dimension of our system and $a$ is suitably chosen to introduce a diffusion process (see next section). Next, we advect the $N_j$ individuals in continuous space using the external velocity (if present). After this step, some of the original $N_j$ individuals have been moved to different regions of space, i.e., to a different box of size $\Delta x$, changing the number of individuals of the new box.  Once we complete the diffusion and advection for all sites, independently one from another, we apply the birth-death processes stochastically  according to the prescribed rates. Note that we do not need to remember the exact position in each site from one step to another: it is enough to know how many individuals of each species  are present at the prescribed site.  In this way, we can efficiently work with an extremely large number of  individuals per site without managing the position of each individual.   This is actually the reason why we can achieve a significant increase in the computational performance. 

The paper is organized as follows. In Sec. \hyperlink{met}{II}, we discuss the details of our method. 
We present a systematic comparison of our approach against known analytical and numerical results in one dimension (Sec. \hyperlink{num}{III}) and two dimensions (Sec. \hyperlink{numdue}{IV}). In Sec. \hyperlink{weak}{V}, this approach is used to extend the previous findings of ~\cite{plummer2018fixation}.  In particular, we investigate the fixation probability of an advantageous species in a two-dimensional weak compressible flow. 

%%%%%%%%%%%%%%%%%%%%%%%%%%%%%%%%
%      
% METHOD
%.    
%%%%%%%%%%%%%%%%%%%%%%%%%%%%%%%%

\section{ II. Method}
\hypertarget{met}{}

The computational approach is described in this section, where, for simplicity, we start by considering a one-dimensional (1D) system with periodic boundary conditions. 
Let $L$ be the size of the 1D lattice, which we discretize with $n$ intervals of size $\Delta x = L/n$.
Each interval $i=1, ....n$ spans the region $x \in [(i-1) \Delta x, i \Delta x]$.
We denote by $N_{i}^{(\beta)}$ the number of individuals in the interval $i$, where $\beta = A,B$ refers, in this case, to the two possible species
(for different realizations, the number of the species may also be greater). At equilibrium and with no external flow, we can define the density $N_0$ of individuals per mesh point corresponding to the overall carrying capacity $N_0 L /\Delta x$. With this definition, we can also think of $N_0$ as the average carrying capacity for a mesh site.
%We denote by $N_0$ the maximum number of individuals for each box of size $\Delta x$. 

At time $t$, our knowledge is given by the set of numbers $N_{i}^{(\beta)}$ for $i=1, ...,n.$ Our task is to compute the evolution of the system at time $t+\Delta t$, where $\Delta t$ represents our time step.

We implement the evolution using four different steps. 
In step 1, we implement a Markov chain with next-neighbor hopping  and periodic boundary conditions, which is known to be consistent with the diffusion equation with  diffusivity $D$  once the hopping probability is given by the relation
\begin{equation}
p \equiv \frac{D \Delta t}{\Delta x^{2} };
\label{eq:p}
\end{equation}
with  $p \ll 1$. 

{\it Step 1}. $\it{Diffusion}$.
For each interval $i$, we compute the particle positions $x_{a}(i) \, (a=1, ...,N_{i})$ according to the rule:
\begin{equation}
x_{a}(i)= \left(i-\frac{1}{2}\right) \Delta x  + \Delta x \bigg(\eta - \frac{1}{2}\bigg)(1+2p),
\label{eq:particle}
\end{equation}
where $\eta$ is a random number that is uniformly distributed $ [0,1]$. In this step, only a small fraction of the $N$ individuals is spread outside of the initial site $i$. Note that we do not assume any knowledge of 
the previous position of the individuals. 

{\it Step 2}. $\it{Advection}$.
Once step 1 is performed, we can compute the advection and obtain
\begin{equation}
x_{a}(i,t+\Delta t)= x_{a}(i)+ u(x_{a}(i))\Delta t
\end{equation}
where $u(x, t)$ is a prescribed advecting field.

{\it Step 3}. $\it{Relabeling}$.
For each off-mesh particle $a$, we can now determine the deme index,
\begin{equation}
j \equiv \left\lfloor \frac{x_{a}(i, t+\Delta t)}{\Delta x}\right\rfloor +1
\end{equation}
and therefore apply the rule
\begin{equation}
{\tilde{N}}_{j}={\tilde{N}}_{j}+1
\end{equation}
to increment the deme occupancy number.
Note that before implementing Eq. (\ref{eq:particle}) we put ${\tilde{N}}_{j} = 0$ for all demes $j=1, ....,n$.
Since from Step 1 to Step 3 we repeat the same operation for both species in the sections, we ignore the label $\beta$ for the different species.

{\it Step 4.} $\it{Birth\, and\, death\, processes}$.
After running step 1 to step 3 for all the intervals, we apply the last step where we execute the rules for stochastic population dynamics for each segment $j$. At this stage, for every $j$ interval, we compute the birth-death process $\tilde{N_j}$ times according to the following rules:
\begin{eqnarray}
{\tilde{N}}_{j}^{(A)}  &=& {\tilde{N}}_{j}^{(A)} +1 \qquad \text{at} \,\, \text{rate} \,\, r_b(A) \\
{\tilde{N}}_{j}^{(A)}  &=& {\tilde{N}}_{j}^{(A)} - 1 \qquad \text{at} \,\, \text{rate} \,\, r_d(A) \\
r_b(A) &=& \mu \Delta t \nonumber \\
r_d(A) &=&  \mu \Delta t \frac{{\tilde{N}}_{j}^{(A)}-1+{\tilde{N}}_{j}^{(B)}(1-s)}{N_0} \nonumber \\
{\tilde{N}}_{j}^{(B)}  &=& {\tilde{N}}_{j}^{(B)} +1 \qquad \text{at} \,\, \text{rate} \,\, r_b(B) \\
{\tilde{N}}_{j}^{(B)}  &=& {\tilde{N}}_{j}^{(B)} - 1 \qquad \text{at} \,\, \text{rate} \,\, r_d(B) \\
r_b(B) &=& \mu \Delta t \nonumber \\
r_d(B) &=&  \mu \Delta t \frac{{\tilde{N}}_{j}^{(B)}-1+{\tilde{N}}_{j}^{(A)}(1+s)}{N_0} \nonumber 
\end{eqnarray}  

where $s$ is the selective advantage, $s > 0$, or disadvantage, $s < 0$, of individuals $A$ with respect to $B$.
Here, $r_b$ and $r_d$ denote the $birth$ and the $death$ probability, respectively.
Note that for each mesh site, the probability to obtain $k$ new offspring or deaths is binomial and it approximates a Poisson distribution
only when the number of individuals considered in the specific process is large enough. This is never the case near the edge of a propagating front and/or
near extinction even for large value of $N_0$.

At the end of step 4, we can put ${N}_{j}^{(\alpha)}={\tilde{N}}_{j}^{(\alpha)}$ and we can start with a new time step.

Let us now briefly comment about our method.  The effect of advection does not change the number of particles, i.e., it is conservative. 
Thus neglecting, for the time being, the death-birth process,
we obtain the equation for each species,
\begin{equation}
\partial_{t} N(x,t)+\partial_{x} (u(x,t)N(x,t)) = D \Delta N.
\end{equation}
The birth-death process is the same one implemented in Ref. ~\cite{plummer2018fixation}. 

On the other side, ignoring diffusion and advection
and neglecting terms of the order of $s/N_0$
inside the noise term ~\cite{pigolotti2013growth}, 
step 4 gives
\begin{eqnarray}
\frac{dN_{A}(t)}{dt} &=& \mu N_{A} \bigg( 1- \frac{N_{A}+N_{B}}{N_0} \bigg)+ \mu s \frac{N_{A}N_{B}}{N_0}\\
&+& \sqrt{N_{A} \mu \bigg(1+ \frac{N_{A} + N_{B}}{N_0}\bigg)} \eta_{A}(t) \nonumber  \\
\frac{dN_{B}(t)}{dt} &=& \mu N_{B} \bigg(1- \frac{N_{A}+N_{B}}{N_0}\bigg)- \mu s \frac{N_{A}N_{B}}{N_0}\\
&+& \sqrt{N_{B} \mu \bigg(1+ \frac{N_{A} + N_{B}}{N_0}\bigg)} \eta_{B}(t)\nonumber 
\end{eqnarray}
where  $\eta_{A}$ and $\eta_{B}$ are independent $\delta$ correlated in time Wiener processes.
Upon defining $c_{A}=N_{A}/N_0$ and $c_{B}=N_{B}/N_0$ and introducing the advection the final equations of motion read:
\begin{eqnarray}
&\partial_{t}c_{A} + \partial_{x} (u c_{A}) =D \Delta c_{A} + \mu c_{A} (1-c_{A}-c_{B}) \\
&+ s \mu c_{A}c_{B}+ \sqrt{\mu \frac{c_{A}}{N_0}(1+c_{A}+c_{B})} \eta_{A} (x,t)  \nonumber \\
&\partial_{t}c_{B} + \partial_{x} (u c_{B}) =D \Delta c_{B} + \mu c_{B} (1-c_{A}-c_{B}) \\
&- s \mu c_{A}c_{B}+ \sqrt{\mu \frac{c_{B}}{N_0}(1+c_{A}+c_{B})} \eta_{B} (x,t)  \nonumber 
\end{eqnarray}
%where $\eta_{A}$ and $\eta_{B}$ are now independent delta correlated in both space and time Wiener processes.
Finally, assuming that $c_{T} \equiv c_{A}+c_{B} \sim 1$ everywhere and upon denoting $f = c_{A}/c_{T}$, we obtain
\begin{equation}
\label{eq_f}
\partial_{t} f + u (x,t) \partial_{x} f = D \Delta f + s \mu f (1-f) + \sqrt{\frac{2 \mu f(1-f)}{N_0}}\eta(x,t).
\end{equation}
We remark that the statistical properties of the system are invariant upon the scaling:
 $\mu \rightarrow 1$, $D \rightarrow \frac{D}{\mu}$, $t \rightarrow t \mu$,
 which is equivalent to working in units  of generation time.

%%%%%%%%%%%%%%%%%%%%%%%%%%%%%%%% 
%
% SUBSECTION - TEST 1 DIMENSION 
% 
%%%%%%%%%%%%%%%%%%%%%%%%%%%%%%%%  

\section{III. Numerical test in one dimension}
\hypertarget{num}{}
In this section, we introduce some numerical tests confined to one-dimensional systems. First, we need to solve the Fisher-Kolmogorov-Petrovsky-Piscounov (FKPP) equation that describes the space-time evolution of a population in a reaction-diffusion system;
in one space dimension, it reads
\begin{equation}
\partial_{t} c = D \partial_{xx} c + \mu c (1-c),
\label{eq:1dimfkpp}
\end{equation}
where $c(x,t)$ is a continuous variable that identifies the concentration of individuals, $D$ is the diffusion coefficient, and $\mu$ is the growth rate.
The uniform solutions of Eq. (\ref{eq:1dimfkpp}) are  $c=1$ and $c=0$ for a stable and unstable state, respectively. In 1995, Mueller and Sowers ~\cite{mueller1995random} 
showed that for $\mu > 0$, the traveling wave solutions to Eq. (\ref{eq:1dimfkpp}) are always characterized by a $\it{compact\, support\, property}$. We can set up initial conditions that depend on $c$ as follows: $c(x,0)\rightarrow1$ as $x \rightarrow -\infty$ and $c(x,0)\rightarrow0$ as $x \rightarrow +\infty$.
For this kind of boundary conditions, we can find a continuous family $\omega$ of traveling wave solutions of the form
\begin{equation} 
c(x,t)= \omega (x-v t),
\end{equation}

where $v$ is the velocity of the traveling wave and $\omega$ is a function that must satisfy the following ordinary differential equation
\begin{eqnarray} 
D \omega''+ v\, \omega'+ \mu \, \omega (1-\omega) = 0, \\
\text{with} \,\, \text{conditions} \, \qquad \omega(- \infty) = 1, \quad \omega (\infty) = 0.  \nonumber 
\end{eqnarray} 

Around the unstable state, $c(x,0)$, the velocity of the front approaches the deterministic continuum minimum value $v_{\text{min}} = 2 \sqrt{D \mu}$. The fronts at this minimum speed are called ``pulled fronts'', which are pulled along by the growth and spreading of small perturbations in the leading edge where $c\ll1$.
We therefore expect this velocity to change due to the discreteness of our model: we are in the presence of a discrete process  in both time and space and the observed value for the Fisher wave velocity propagation is lower than the deterministic one. Brunet and Derrida ~\cite{brunet} 
gave an estimation of how far the Fisher wave value has to be 
from the continuum wave speed as
\begin{equation}
v \sim \sqrt{D \mu} \bigg[ 2- \frac{\pi^{2}}{(\ln N)^{2}} \bigg].
\label{eq:wavespeed}
\end{equation}
From Eq. (\ref{eq:wavespeed}), one can clearly observe that the convergence to the continuum limit is extremely slow as  $N \rightarrow \infty$. Fluctuations have been considered by the Doering, {\it et al.} conjecture ~\cite{doering2003interacting} 
adding a noise term to the FKPP equation;
for the strong noise regime (or weak growth limit), they found that the speed value goes according to
\begin{equation}
v \sim D \mu N.
\end{equation}

%%%%%%%%%%%%%%%%%%%%%%%%%%%%%%%% 
%
% Figura_convergenza_mu_10 
%
%%%%%%%%%%%%%%%%%%%%%%%%%%%%%%%%
   
\begin{figure}[!ht]
\includegraphics[width=\linewidth]{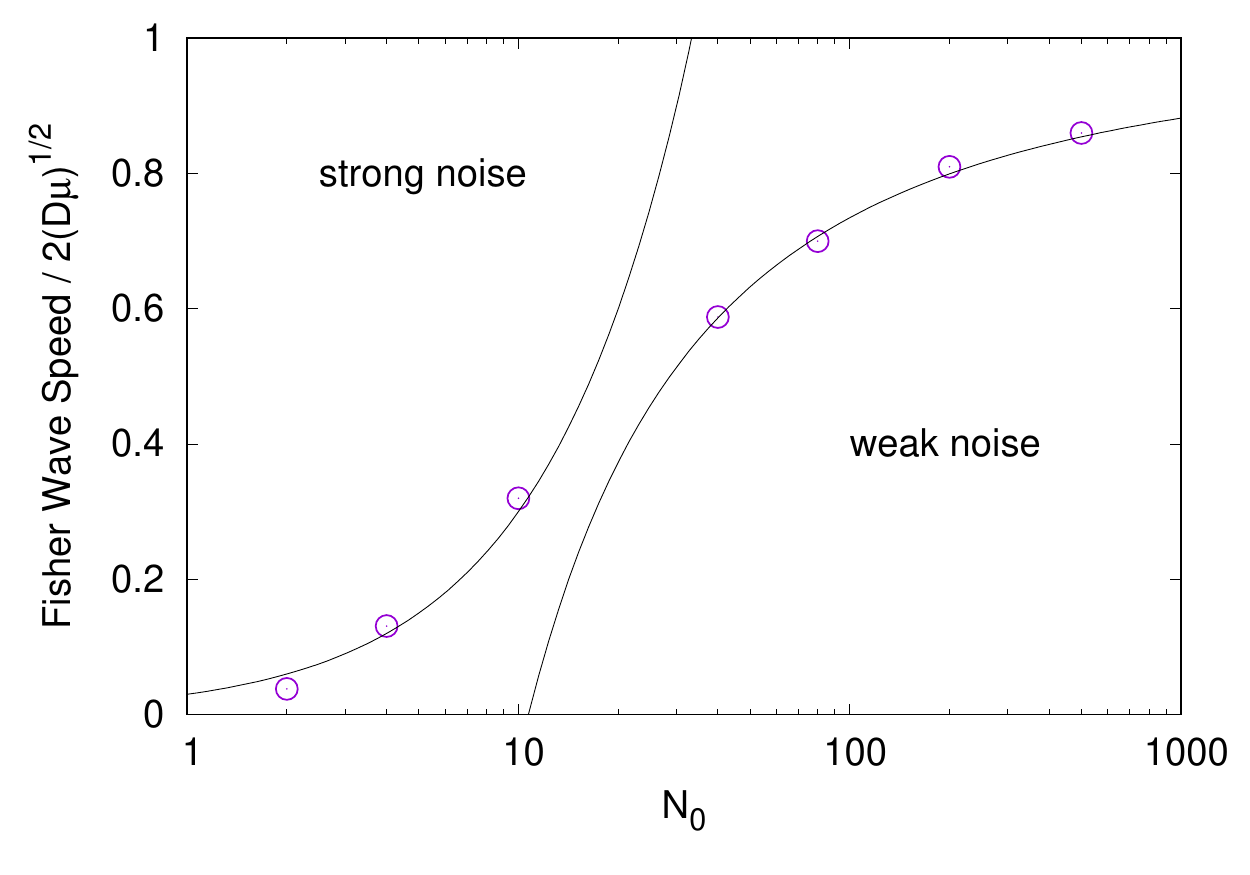}
\caption{Algorithm convergence tests: Fisher wave behavior varies with the number of individuals per site. It is possible to distinguish two theoretical limits: on the left, the strong noise trend and, on the right, the weak noise one. We performed simulations for $\mu = 10$ and $D = 0.001$; the circles indicate the results of our simulations that are, asymptotically, in very good agreement with the theoretical lines.}  
\label{fig:figura_convergenza_mu_10.eps}
\end{figure}

In Fig. ~\ref{fig:figura_convergenza_mu_10.eps} the normalized Fisher wave speed versus the number of individuals per site $N_0$ is shown. There are two theoretical estimates, corresponding to the weak and strong limits. The simulations, identified by dots, are consistent with the theoretical lines: with $10$ particles per site we are in the strong noise regime, where the Fisher velocity is equal to $\approx$ 0.3 times the theoretical expectation.
In this work, simulations are performed in the weak noise regime, where the velocity of the genetic wave $v_{g}$ is $\simeq v_{\text{min}}$.

%%%%%%%%%%%%%%%%%%%%%%%%%%%%%%%%
%
% FIGURA Heterozigosity
%
%%%%%%%%%%%%%%%%%%%%%%%%%%%%%%%%  

\begin{figure}[!htbp]
\includegraphics[width=\linewidth]{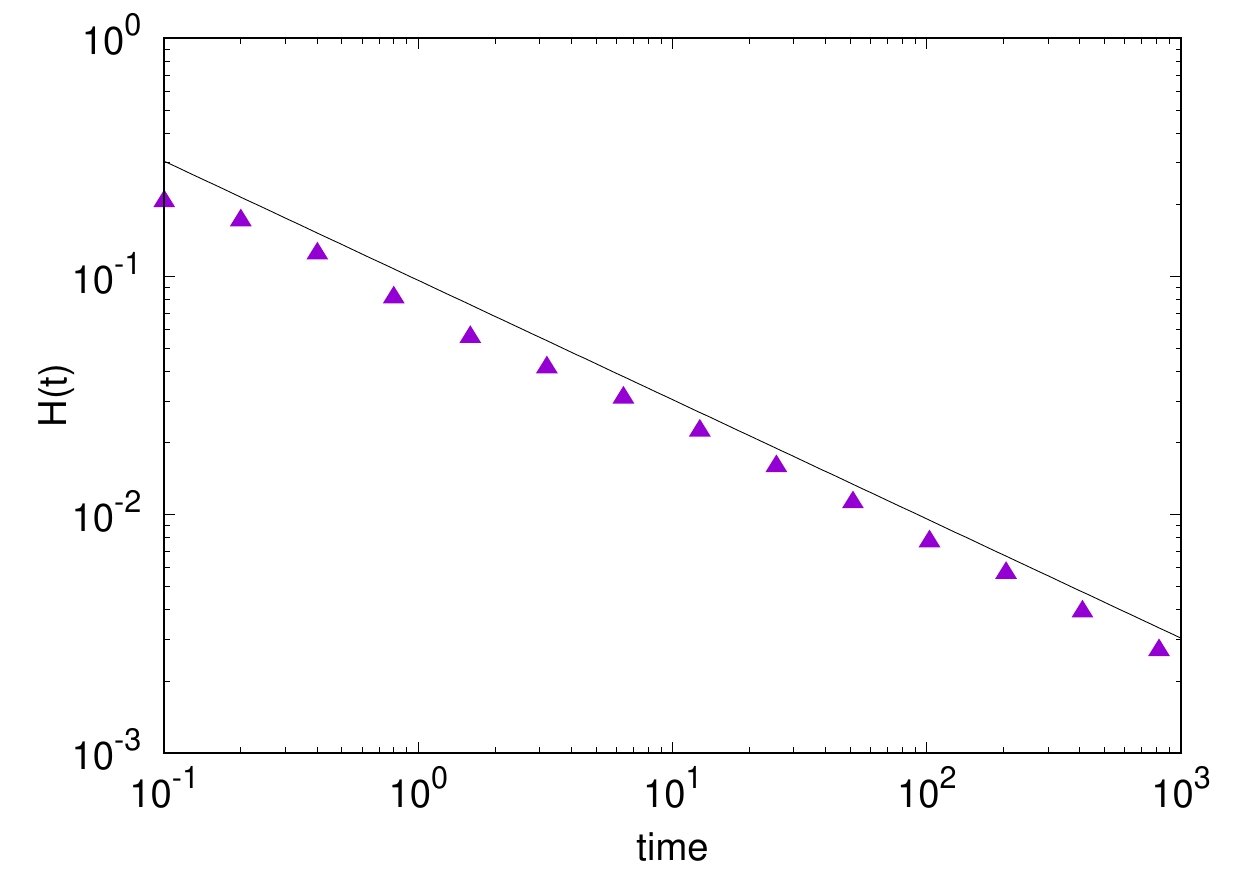}
\caption{Log-log plot of the decay of one-dimensional local heterozygosity, $ H(t) $, as a function of time. The black continuous line shows the theoretical heterozygosity in 1D, $t^{-1/2}$, and the purple symbols show our simulations. The error bar is calculated on 500 cases and the variance is smaller than the symbol size.} %performed on a  $256 \times 256$ lattice with $20$ individual per box.}
\label{fig:figure_heterozigosity.eps}
\end{figure}

The diversity of a population composed of two genotypes in one dimension is measured by the heterozygosity ~\cite{korolev2010genetic},
\begin{equation}
H(x,x'; t) \equiv \langle  f(x,t) [1-f(x',t)] \rangle.
\end{equation}
This quantity is given by the product of the two fractions $f(x,t)$ and $(1-f(x',t))$ and it defines the probability that two selected individuals, chosen at random, are from different species (carry different alleles) ~\cite{pigolotti2013growth}. For homogeneous conditions, $H(x,x';t)$ depends on the $r=|x-x'|$.
The heterozygosity becomes zero when  there is fixation of one of the two genotypes. 
Moreover, it is known that in a one-dimensional system, $H(t) \equiv H(x,x;t)$  decays  in time as $t^{-1/2}$. In  Fig. ~\ref{fig:figure_heterozigosity.eps}, we have tested this theoretical prediction using our methods with $N_0=50$ on a domain with periodic boundary conditions discretized with 512 mesh points: the result very clearly confirms the theoretical behavior.

%%%%%%%%%%%%%%%%%%%%%%%%%%%%%%%%
%
% FIGURA ADVANTAGE KIMURA 1D no velocity
% 
%%%%%%%%%%%%%%%%%%%%%%%%%%%%%%%% 
Next,  to further validate the algorithm, we calculate, in the absence of advection, 
the fixation probability  given by Eq. (\ref{eq:kim}). 

\begin{figure}[!htbp]
\includegraphics[width=\linewidth]{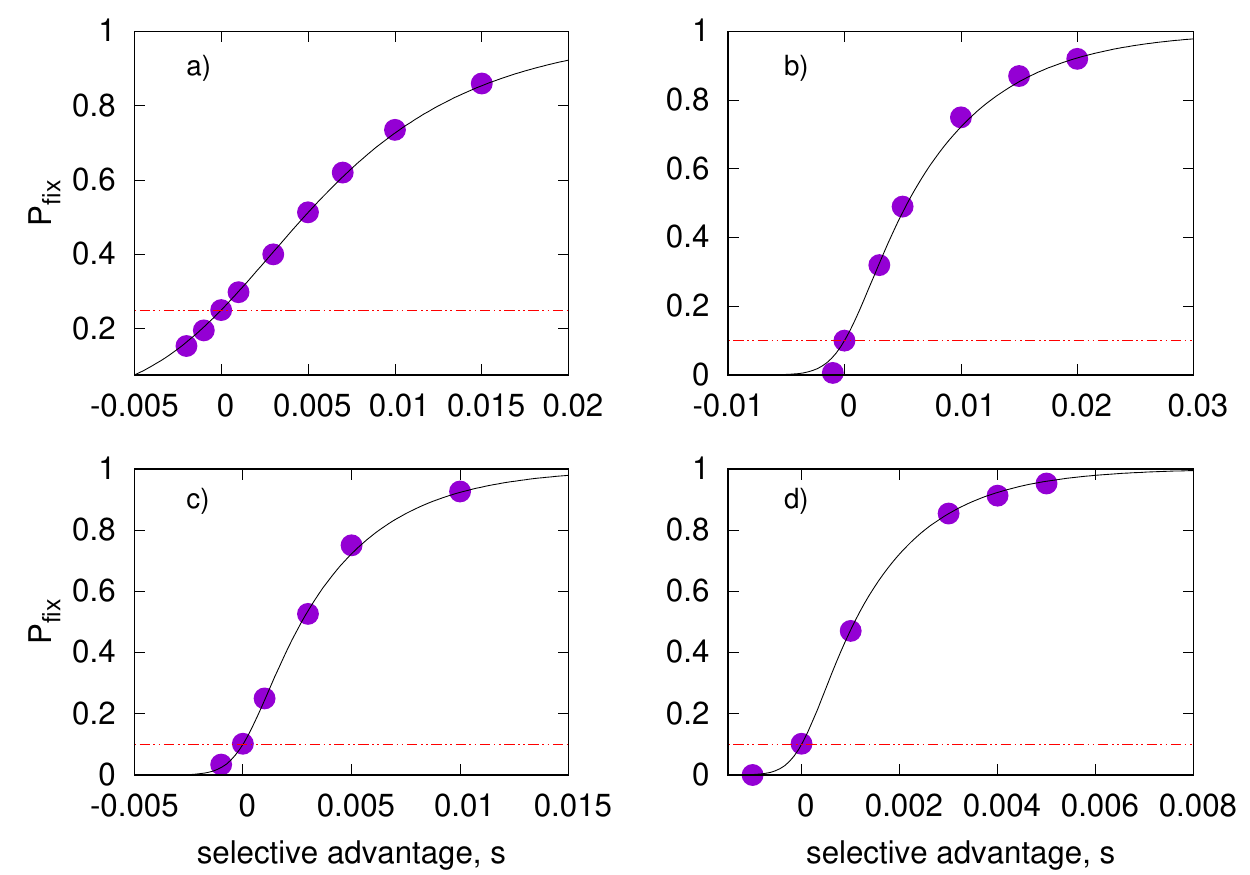}
\caption{Fixation probability of one species, in a one-dimensional domain, vs different values of the selective advantage. In each plot, the Kimura formula is reported (black solid line) by fixing $N_{0}$, the number of particles per site, and $f_{0}$, the initial fraction of a species (red dashed line).  Our no-flow results for the fixation probability are illustrated with the solid circles; the lines and the results are in very good agreement for every case. The length of the domain of size $L = 2 \pi$ is divided into $128$ intervals. The values of $N_{0}$ and $f_{0}$ for each graph are  (a)   $N_{0} = 4$, $f_{0} = 0.25$; (b) $N_{0} = 10$, $f_{0} = 0.1$; (c) $N_{0} = 20$, $f_{0} = 0.1$; (d) $N_{0} = 50$, $f_{0} = 0.1$. $N$ in Eq. (\ref{eq:kim}) is $128\, N_0$.}
 \label{fig7}  
 \end{figure}

In Fig. \ref{fig7}, different panels corresponding to a different number of individuals per box are shown. In our simulations, we focus on small selective advantages, in order to study more realistic cases.
Our results are in good agreement with the theoretical predictions (continuous black lines, in the figures).

\section{IV. Numerical test in  two dimensions}
\begin{figure}[!htbp]
\includegraphics[width=\linewidth]{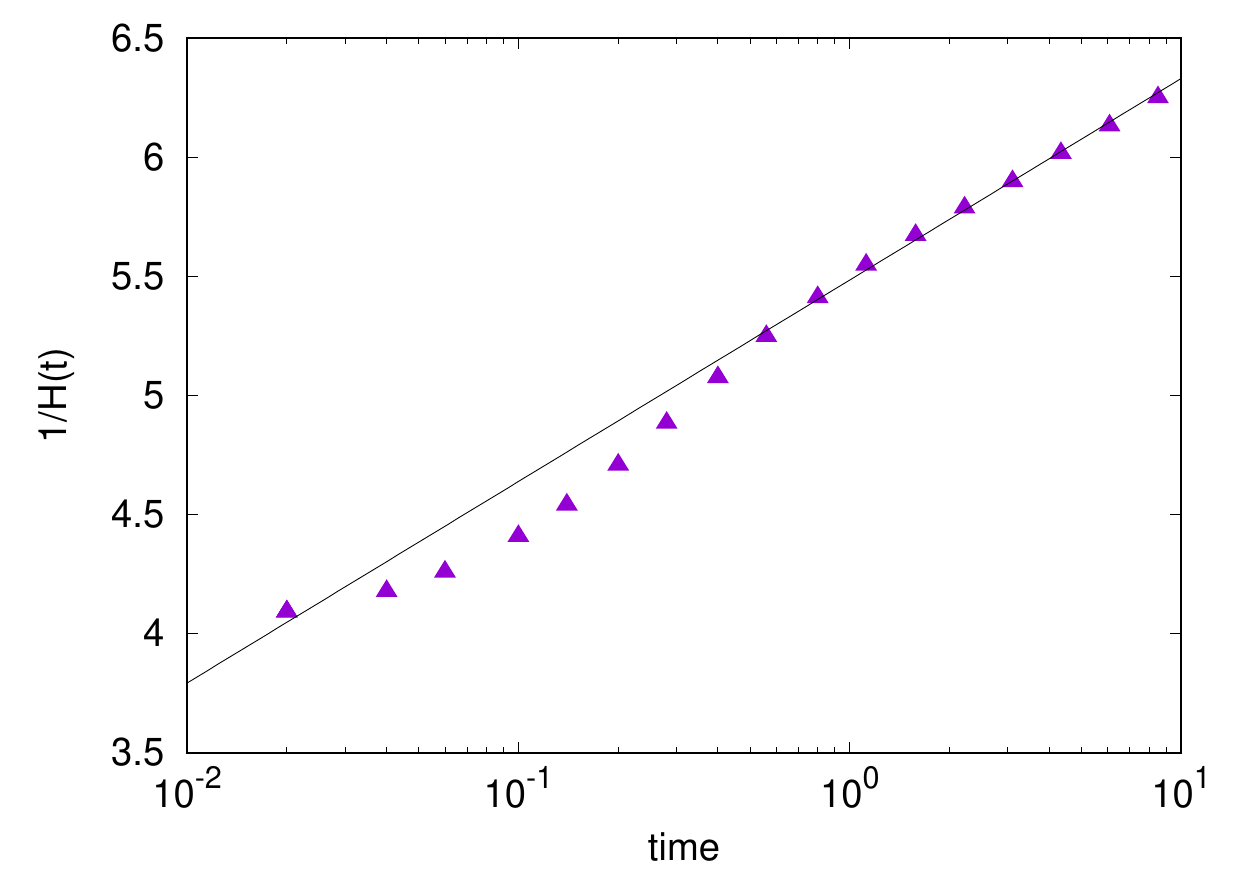}
\caption{Behavior in log-linear scale of the inverse of the heterozygosity, $1 / H(t) $, as a function of time in 2D. The symbols, representing the results of our simulation results, are in good agreement with the black solid line that indicates the theoretical trend, $\frac{1}{H(t)} \sim \ln(t)$.}

\label{fig:figura9.eps}  
\end{figure}

\hypertarget{numdue}{}
In this section, we implement the method previously introduced (Sec. \hyperlink{met}{II}) and validated for a one-dimensional system on a two-dimensional configuration. Following the same schematic procedure of the 1D case we start by estimating the heterozygosity parameter. 
It is known that in two spatial dimensions, the local heterozygosity decay in time is slower compared to 1D: it goes to zero as $H(t) \sim 1 / \ln(t)$ ~\cite{korolev2010genetic, pigolotti2013growth}. To check whether our method is able to exhibit such (slow) decay, we specifically perform a set of numerical simulations with $N_0=20$ on a domain with periodic boundary conditions and $256^2$ mesh point.
In Fig. ~\ref{fig:figura9.eps}, such slow logarithmic decay is appreciable. In this figure, we plot $1/H(t)$ versus time. Note that starting with well mixed conditions,  $H(0)=1/4$. Therefore, $1/H(t)$ is $4$ at $t=0$ and grows in time as $\ln(t)$, as shown in the figure. The loss of the genetic variability given by our simulations (purple triangles) is in agreement with the theory (black solid line).

\begin{figure}[!htbp] 
\includegraphics[width=\linewidth]{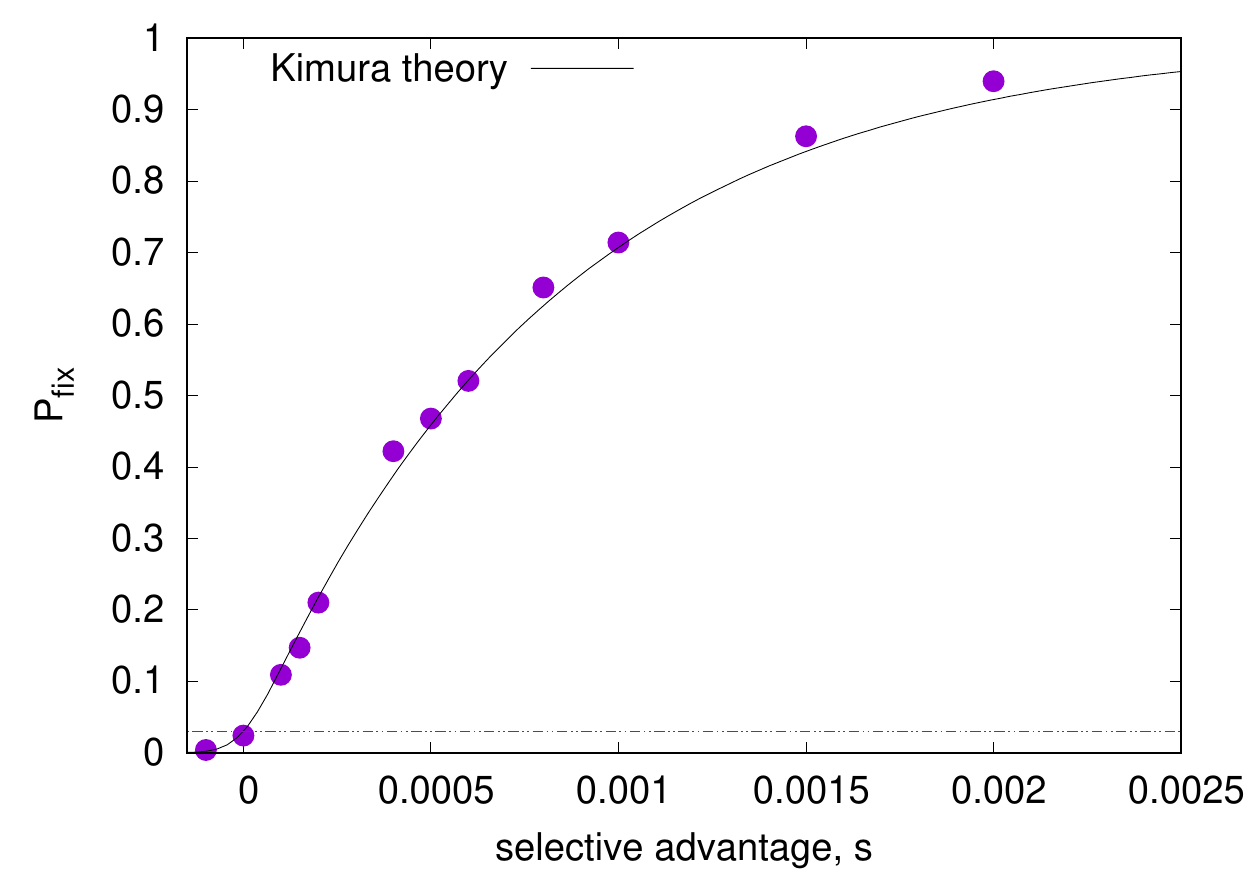}
\caption{Probability of fixation as a function of the selective advantage in the absence of advection. Simulations are performed on a $64 \times 64$ lattice with, initially, $10$ individuals per box, a diffusion coefficient of $D=0.01$, and an initial fraction 
of $f_{0} =0.03$ (horizontal red dashed line). 
Our no-flow results for the fixation probability  are illustrated with the solid circles and the theoretical prediction of Kimura by the solid black line. 
}
\label{fig:figura2d}
\end{figure} 
The second step, as in the numerical validations in 1D, is to verify Kimura's formula, given by Eq. (\ref{eq:kim}), for the two-dimensional system in the absence of advection.
The formula for the probability of fixation is still valid for higher dimensions and our results together with the theoretical prediction (solid black line) show an unequivocal agreement in Fig. ~\ref{fig:figura2d}. 

%%%%%%%%%%%%%%%%%%%%%%%% 
%
% CASES WITH VELOCITY 2D
%
%%%%%%%%%%%%%%%%%%%%%%%% 
 
\section{V.  Weak compressible flow in D=2}
\hypertarget{weak}{}
Before adding an advecting velocity field to our two-dimensional system, we briefly discuss the main results achieved by
Plummer {\it et al.}  ~\cite{plummer2018fixation}, where a particular configuration of the velocity field was used, given by
\begin{equation}
\label{ux}
u(x)=u_{0} \sin\bigg(x-\frac{\pi}{2}\bigg).
\end{equation}
For small enough $u_0$, the flow field in (\ref{ux}) is weakly compressible, i.e., the condition $c_A+c_B=1$ is valid within a small percentage (up to $4$ percent for $u_0=0.05$ on a domain size $2\pi$). We will test whether, as in 1D, the Kimura formula is still valid  provided we define $N$ as an effective population size, $N_{\text{eff}}$.
For $s\rightarrow 0$, it has been shown in   ~\cite{plummer2018fixation} that $N_{\text{eff}}$ depends only on the diffusion constant $D$,  $u_0$, and on the maximum number $N_0$  of individuals per site. The crucial point is to recognize that near to the source, one can define a characteristic scale, $l_s = \sqrt{D/u_0} $. Any organism that moves significantly farther than $l_s$ from the source is unlikely to be able to return and has, therefore, a negligible chance of fixation as it is drawn into the sink. It follows  (see  ~\cite{plummer2018fixation} for details) that $N_{\text{eff}}$ can be estimated as
\begin{equation}
N_{\text{eff}}=B_{1} \rho_{0} \sqrt{\frac{D}{u_{0}}} ,
\label{neff}
\end{equation}
where $B_{1}$ is a constant of the order of unity and $\rho_0$ is the density at each point, namely $N_0/ \Delta x$. 

Following \cite{plummer2018fixation}, one simple way to understand the physical meaning of Eq. (\ref{neff}) is to consider the deterministic case, i.e., Eq. (\ref{eq_f}) in the limit $N_0 \rightarrow \infty$, and assume an initial population $f=1$ in a small box $\Delta$ at the location $x_0$, and zero otherwise. 
Then, the population, whose spatially averaged initial ratio is $\bar f_0 \equiv \langle f_0 \rangle= \Delta/L$, where $ L=2 \pi$, evolves to an asymptotic value, $f_{\infty}(x_0) = lim_{t \rightarrow \infty} f_t(x_0) $, which depends on $x_0$. The ratio $f_{\infty}(x_0)/\bar f_0$ is a function of $x_0$  and shows a Gaussian-like behavior in terms of $x_0-x_s$, where $x_s$ is the position of the source with a variance proportional to $l_s$ and $f_{\infty}(x_s)/\bar f_0 \gg 1$. This is equivalent to saying that for $s=0$, there is a significant advantage for the offspring occurring near the source and a strong disadvantage for those occurring downstream. This implies that the effective population size (for small $s$) is the one corresponding to the population size close to the source, i.e., at distance $l_s$ from the source.   
Following \cite{lieberman}, one simple way to understand this result is to consider a simple toy model on a linear graph where the source is a relative ``cold" site (node of the graph) with respect to the downstream ``hot" sites, where ``cold" and ``hot" refer to the probability for an offspring to be advected by the external flow using the same language of \cite{lieberman}. 

\begin{figure}[!htbp]
\includegraphics[width=\linewidth]{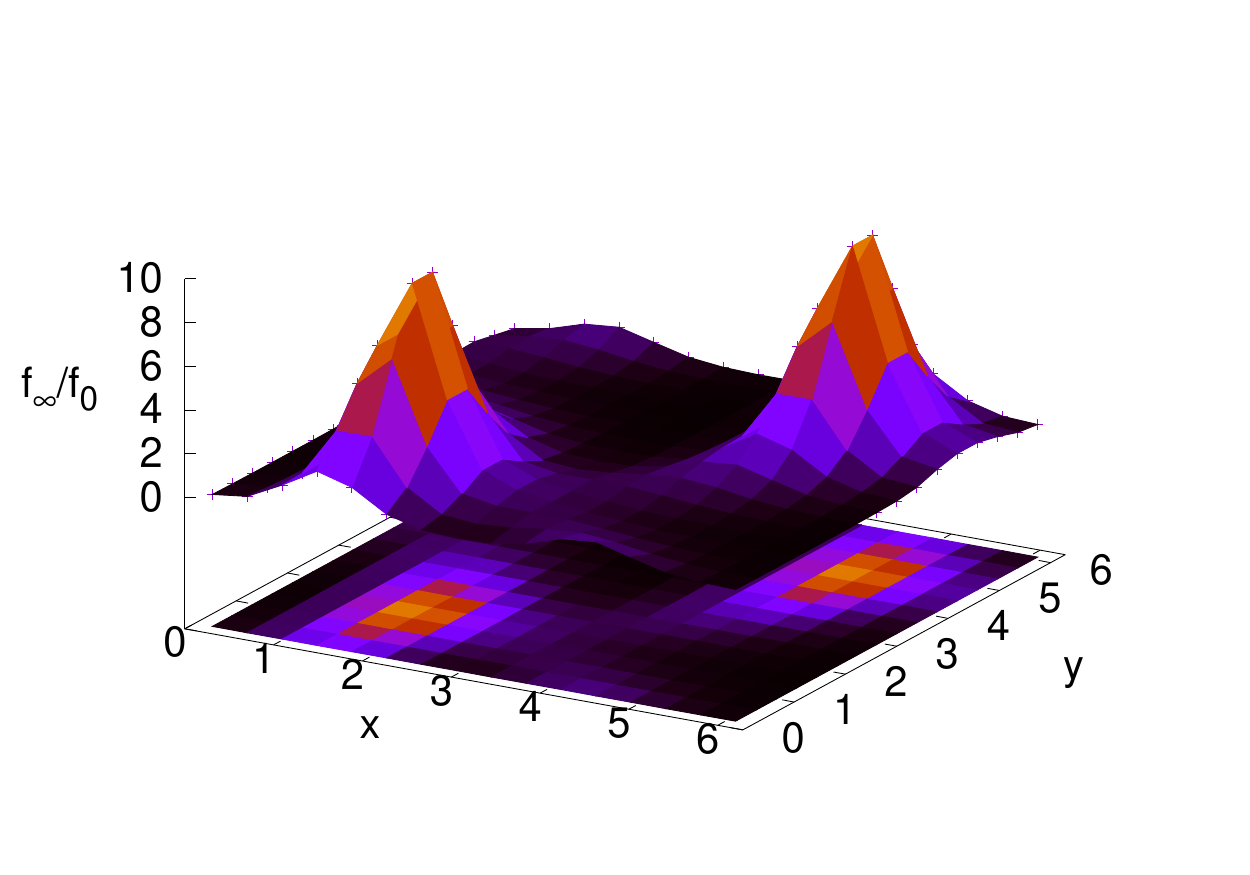}
\caption{Shape of the asymptotic fraction, $f_{\infty}$(x,y), normalized by the initial fraction, $\bar f_0$. Populations starting close to the source become increasingly larger.
}
\label{fig:figura_f_asint.eps}
\end{figure}  

\begin{figure}[!htbp]
\includegraphics[width=\linewidth]{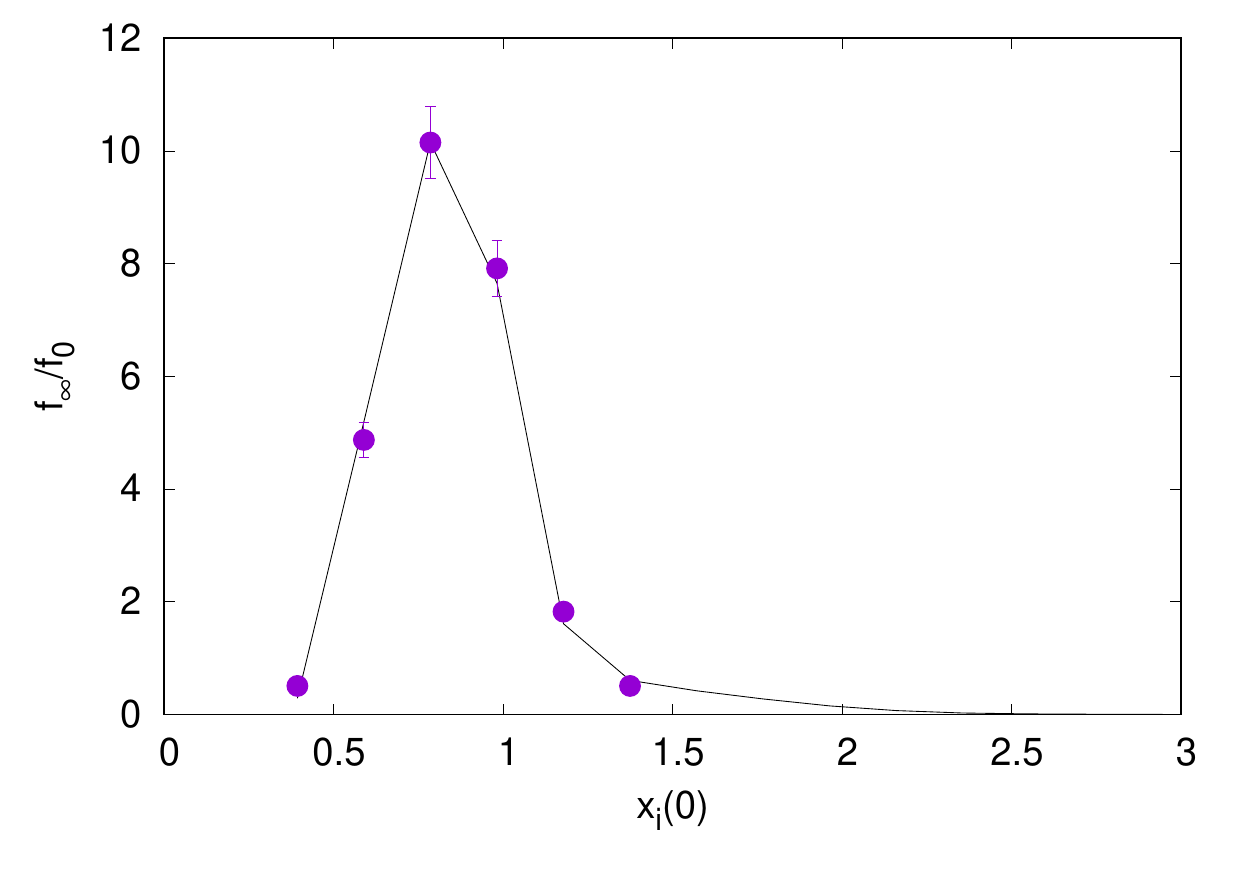}
\caption{Enhancement effect due to the presence of a source.
Comparison between numerical simulation $P_{\text{fix}} / \bar f_0$ (symbols) and the deterministic line $f_{\infty}/ \bar f_0$
 (solid black line). For this simulation we implement 250 cases with a diffusivity of $D = 10^{-2}$, two particles per site with a grid base of $64 \times 64$, and a velocity of $u_{0}=0.05$. The error bar is calculated on 250 cases.}
\label{fig:figura_risultato.eps}
\end{figure}

The same considerations can be made for the two-dimensional version of the same problem. For this purpose, we consider the following flow
\begin{eqnarray}
\label{flow2d}
u(x)&=&u_{0} \sin \bigg(x-\frac{\pi}{2}\bigg)\sin y \\
\nonumber
u(y)&=-&u_{0} \sin \bigg(x-\frac{\pi}{2}\bigg)\cos y
\end{eqnarray}
with periodic boundary conditions and a domain of size $(2\pi)^2$. In Fig. ~\ref{fig:figura_f_asint.eps}, the final fraction of the initially localized species $f_{\infty}(x_0,y_0)/ \bar f_0$ is shown where, now, $\bar f_0 = \Delta^2/(4\pi^2)$. Two peaks are clearly visible in correspondence with the sources, representing the upwelling regions. The asymptotic  value of $f_{\infty}/\bar f_0$ is increasing in proximity of the sources while being it is reduced moving away from them. In  Fig. ~\ref{fig:figura_risultato.eps}, we show,  with a black line, a one-dimensional section (along the $y$ axis) of the two-dimensional behavior of $f_{\infty}/\bar f_0$.  Since for $s=0$ $P_{\text{fix}} = \bar f_0$, one can consider the black line as the increase in $P_{\text{fix}}$ due to the effect of the velocity field near to the source. To validate this interpretation, as well as the quality of our method, we performed a series of numerical simulations with $N_0=2$  at $s=0$ using the same initial conditions of the deterministic simulations. After estimating  the fixation probabilities, we compute the increase of $P_{\text{fix}}$ as a function of the initial position, $x_0$. The results are shown as symbols in  Fig. ~\ref{fig:figura_risultato.eps} where an excellent agreement is visible with the deterministic value of $f_{\infty}/\bar f_0$. This result demonstrates that the mechanism described in  ~\cite{plummer2018fixation}, for small enough $s$ should be true for the two-dimensional flow considered here. 

\begin{figure}[!htbp]
\includegraphics[width=\linewidth]{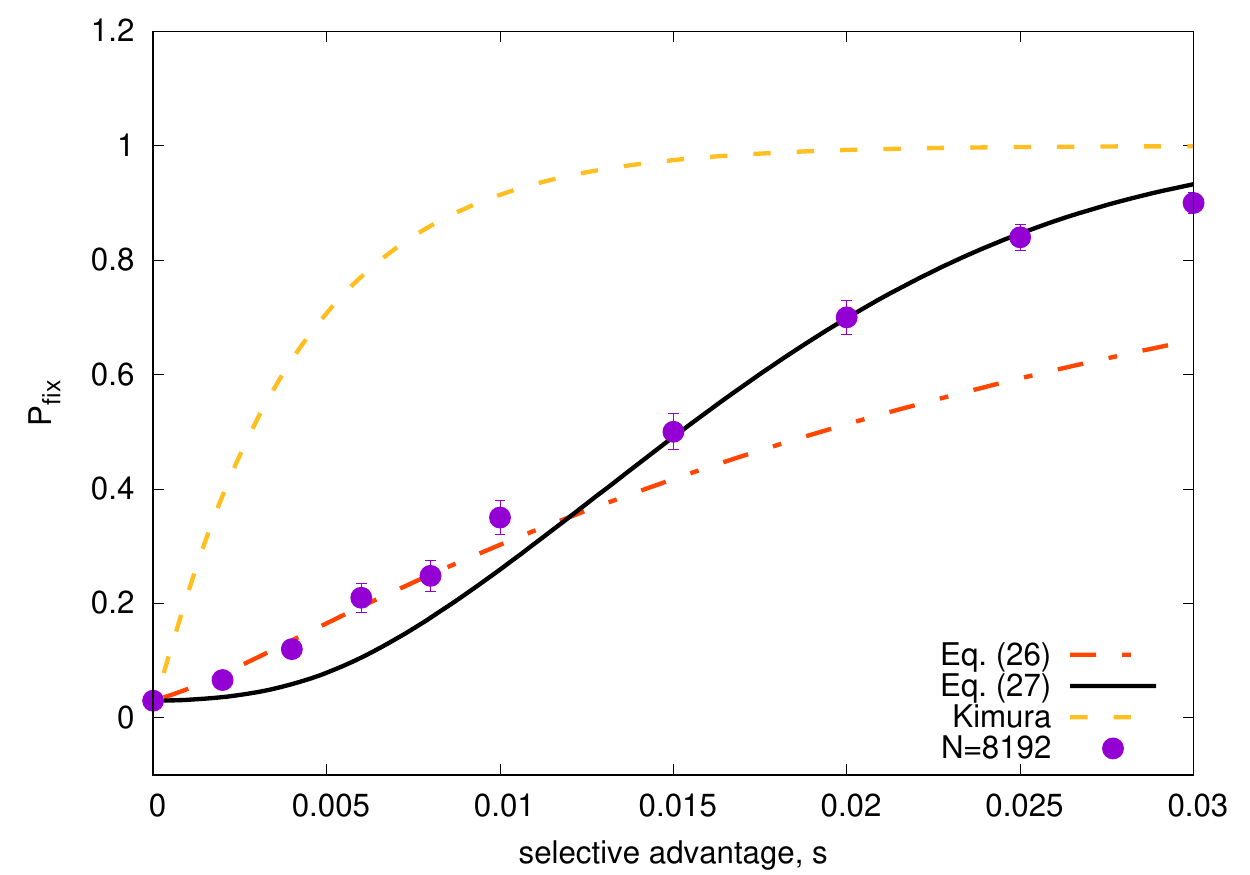}
\caption{Probability of fixation under a 2D velocity field as a function of the selective advantage. The yellow dashed line represents the Kimura theoretical line in the absence of flow that follows Eq. (\ref{eq:kim}).
The base grid is $64 \times 64$ with two individuals per cell, so the total number of individuals is $N=8192$. In addition, the diffusivity parameter is $10^{-2}$ and the velocity value is $u_{0}=0.05$. The black continuous line and the red dot-dashed line are the theoretical predictions and our simulations are illustrated by symbols, with an error bar of about $5 \%$, both in good agreement.}
\label{fig:figura2D_advantage8.eps}
\end{figure}

Based on the previous results, we can generalize Eq. (\ref{neff}) for the two dimensional case as follows:
\begin{equation}
\label{neff2}
N_{\text{eff}} = 2 B_{1}^2 \rho_{0} \frac{D}{u_{0}}
\end{equation}
The factor $2$ in Eq. (\ref{neff2}) comes from the fact that for our flow field, given by Eq. (\ref{flow2d}), we have two sources and two sinks.
Using a grid resolution of $64^2$, with $N_0=2$, we have computed $P_{\text{fix}}$ as a function of $s$ as reported in Fig. ~\ref{fig:figura2D_advantage8.eps}.  Two different behaviors can be observed depending on the value of $s$. The small $s$ region is very
well fitted by the Kimura formula (\ref{eq:kim}) with an effective population size given by Eq. (\ref{neff2}) and with the same value of $B_1 = 3.5$ used in ~\cite{plummer2018fixation}.

From Fig. ~\ref{fig:figura2D_advantage8.eps}, it is clear that the behavior of $P_{\text{fix}}$, for large enough $s$, is controlled by a different value of the effective populations size, hereafter referred to as $N_g$.  In one dimension, following ~\cite{plummer2018fixation},  the effective population size is estimated by considering the scale $\delta$ near to a source in $x_s$ where $u_0 \delta \sim B_2 2\sqrt{D\mu s}$, with $B_2$  another constant of the order  of $1$: an initial population in $ x \in [x_s - \delta, x_s+\delta]$ can develop a Fisher genetic wave at speed $2 \sqrt{D \mu s}$, which is supported by the velocity field. Only Fisher genetic waves that start in this interval are able to cross the system; this provides an estimate  $N_g = 2 \delta(s) \rho_0$. In two dimensions, the same argument gives:
\begin{equation}
\label{neff_sbig2d}
N_g \sim 4 \delta^2 \rho_0= 4  \left[ \frac{2\sqrt{D\mu s}}{u_0} \right]^2 \rho_0.
\end{equation}
Using Eq. (\ref{neff_sbig2d}), we obtain the curve (black) of Fig.  ~\ref{fig:figura2D_advantage8.eps}, which provides an excellent fit of the numerical simulations.  

\begin{figure}[!htbp]
\includegraphics[width=\linewidth]{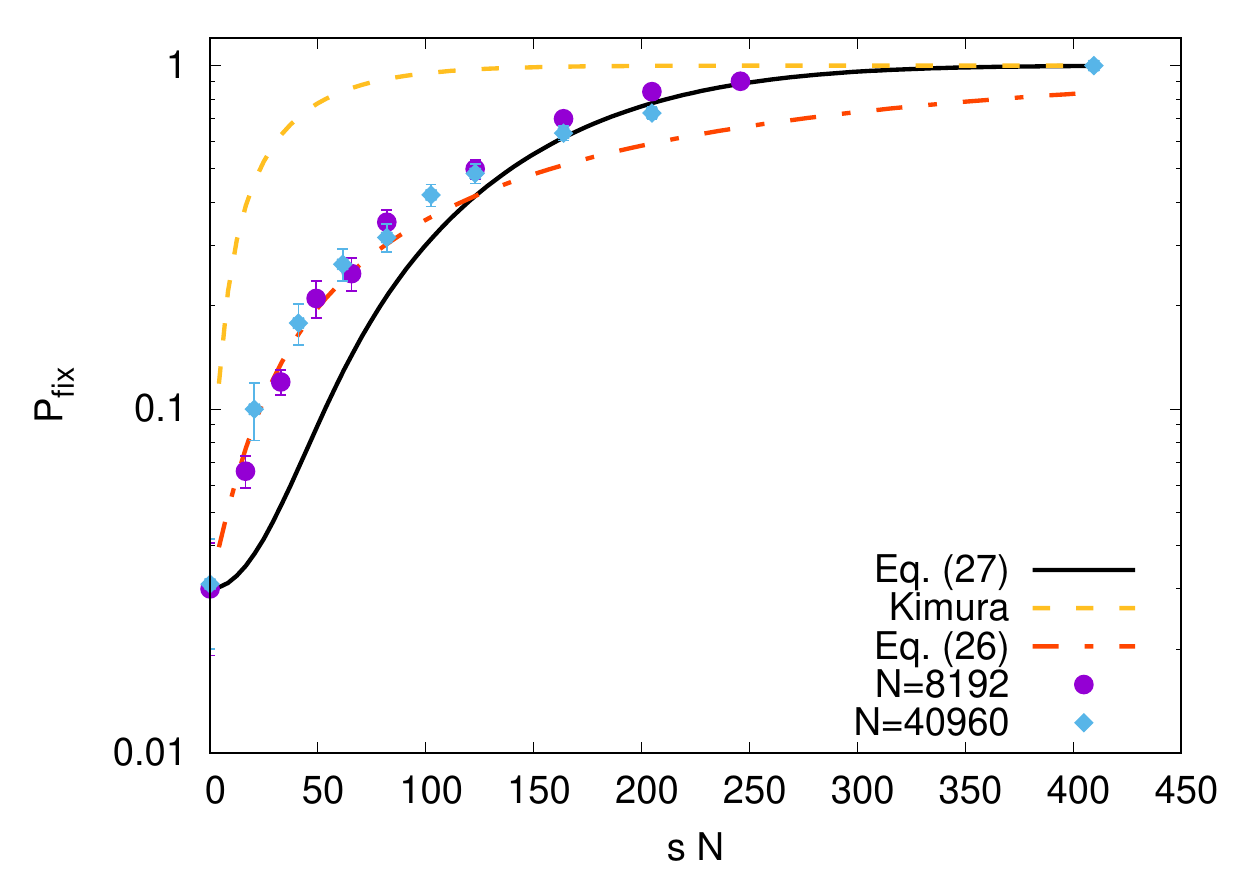}
\caption{Kimura probability of fixation in the presence of the velocity field. Calculations were carried out with an overall mesh size of $64 \times 64$ and with $10$ individuals per cell. The total number of individuals are $N=40960$.
The yellow dashed line shows the theoretical Kimura's trend, while the red dot-dashed and the black continuous lines represent the numerical prediction. Dots and diamonds represent our simulations for two different $N$, 8192 and 40960, respectively, with an error bar about $5 \%$.}
\label{fig:figura2D_advantage4.eps}
\end{figure}

Finally, since both $N_{\text{eff}}$ of Eq. (\ref{neff}) and $N_g$ given by Eq. (\ref{neff_sbig2d}) are proportional to $N =  64^2 N_0$ for our simulations, we can easily predict that upon increasing $N_0$, the fixation probability will follow the same master curve if plotted as a function of $s N$. To demonstrate this and to validate the quality of our method for large $N_0$, we show, in Fig. ~\ref{fig:figura2D_advantage4.eps}, $P_{\text{fix}}$ as obtained for the same flow as Eq. (\ref{flow2d}) for $N_0=2$ and $N_0= 10$. The red dot-dashed and black continuous curves obtained using the prescriptions discussed above for small $s$ and large $s$, respectively, 
provide an excellent fit for the numerical results.  Overall, the results discussed in this section extend the ones previously obtained in ~\cite{plummer2018fixation} and demonstrate the validity of our method for population dynamics advected by an external compressible velocity field. 

\section{VI. Conclusions}
In the present work, we developed a numerical method suitable for accurately and efficiently investigating the behavior of population dynamics and genetics under flow.
This approach allows for the study of a large number of individuals by, first, implementing the diffusion and advection processes, particle by particle, and afterwards, for each box composing the 2D lattice, performing the birth and competition steps.

In order to test and validate our method, we considered a one-dimensional system. We implemented the FKPP equation, analyzing the algorithm convergence. After that, we applied this method to the heterozygosity and Kimura formula and we found a very good agreement between the theoretical and simulated results.  The method we propose does not require any dynamic management of particle positions and has no limitations on the number of individuals for mesh points. Both features  imply  major simplifications in computer coding, especially for a large number of individuals and for parallel computation. It is worth remarking that for a large number of individuals, one can increase the computational efficiency of our method by directly sampling the binomial distribution in each mesh point along the lines discussed in \cite{binomial}.

For the 2D system, we retraced the procedural scheme of the one-dimensional system and we investigated the larger system under an advection field composed of two sinks and two sources.
Our main result was to find, for the 2D system, a net growth of particles born in proximity of a source, as compared to the individuals at different initial positions.

Many interesting studies can follow up on our work.
One of these would be to implement a realistic oceanographic advection field and to understand the population and genetic evolution.
Another topic to investigate could be the study of the effect of stochastic fluctuations in antagonist population dynamics and the exploration of the effect of external velocity on the genetic nucleation theory.
%%%%%%%%%%%%%%%%%%%%%%%%

% Acknowledgments

%%%%%%%%%%%%%%%%%%%%%%%%
\section{Acknowledgments}
The authors would like to thank David Nelson for useful discussions. The work has been performed under the Project HPC-EUROPA3 (Project No. INFRAIA-2016-1-730897), with the support of the EC Research Innovation Action under the H2020 Programme HPC-LEAP; in particular, the authors gratefully acknowledge the computer resources and technical support provided by SurfSARA.
%\newpage
%%%%%%%%%%%%%%%%%%%%%%%%

% Create the reference section using BibTeX:

%%%%%%%%%%%%%%%%%%%%%%%%
%\bibliographystyle{unsrt}
%\bibliography{biblio.bib}

\end{document}